\documentstyle[12pt]{article}

\newcommand{\be}{\begin{equation}}
\newcommand{\ee}{\end{equation}}
\newcommand{\bea}{\begin{eqnarray}}
\newcommand{\eea}{\end{eqnarray}}
\newcommand{\nn}{\nonumber}
\newcommand{\p}[1]{(\ref{#1})}

\begin{document}
\begin{flushright}
hep-th/9508152\\
August 1995
\end{flushright}
\centerline{\large\bf Off-shell (4,4) supersymmetric sigma models
with torsion}
\vspace{0.2cm}
\centerline{\large\bf in harmonic superspace}
\vskip0.6cm
\centerline{{\large Evgenyi A. Ivanov}}
\vskip.3cm
\centerline{\it Bogoliubov Laboratory of Theoretical Physics, JINR,}
\centerline{\it 141 980, Dubna, Russian Federation}
\vskip.5cm

\begin{abstract}{\small
We present a manifestly supersymmetric off-shell
formulation of a wide class of $(4,4)$ $2D$ sigma models
with torsion and non-commuting left and right complex
structures in the harmonic superspace with a double set of
$SU(2)$ harmonic variables. The distinguishing features of the
relevant superfield action are: (i) in general nonabelian and
nonlinear gauge invariance ensuring a correct number of
physical degrees of freedom; (ii) an infinite tower of auxiliary fields.
This action is derived from the most general one by imposing the
integrability condition which follows from the commutativity of the
left and right analyticity-preserving harmonic derivatives.
For a particular class of such models we explicitly demonstrate the
non-commutativity of complex structures on the bosonic target. }
\end{abstract}
\vskip.5cm

\noindent{\bf 1. Introduction.}
Remarkable target geometries of $2D$ sigma models with
extended worldsheet SUSY are
revealed most clearly within manifestly supersymmetric off-shell superfield
formulations of these theories. For torsionless $(2,2)$ and $(4,4)$
sigma models the relevant superfield Lagrangians were found to coincide
with (or to be directly related to) the
fundamental objects underlying the given geometry:
K\"ahler potential in the $(2,2)$ case \cite{Zum},
hyper-K\"ahler or quaternionic-K\"ahler potentials in the flat or
curved $(4,4)$ cases [2 - 5]. One of the basic advantages of such a
description is the possibility to explicitly compute the corresponding
bosonic metrics (K\"ahler, hyper-K\"ahler, quaternionic ...) starting
>from an unconstrained superfield action \cite{{GIOShk},{DVal}}.
To have superfield off-shell formulations with all supersymmetries manifest
is also highly desirable while quantizing these theories. For example, this
simplifies proofs of the ultraviolet finiteness.

An important wide class of $2D$ supersymmetric sigma models
is presented by $(2,2)$ and $(4,4)$ models with torsionful bosonic target
manifolds and two
independent left and right sets of complex structures (see,
e.g. \cite{{GHR},{HoPa}}).
These models and,
in particular, their group manifold WZNW representatives \cite{belg}
can provide non-trivial
backgrounds for
$4D$ superstrings (see, e.g., \cite{Luest}) and be relevant to
$2D$ black holes \cite{RSS}.
A manifestly supersymmetric formulation of $(2,2)$ models with
commuting left and right complex structures in terms of chiral and
twisted chiral $(2,2)$
superfields and an exhaustive discussion of their geometry have been
given in
\cite{GHR}. For $(4,4)$ models with commuting structures there exist
manifestly supersymmetric off-shell formulations
in the projective, ordinary and $SU(2)\times SU(2)$ harmonic $(4,4)$
superspaces [11-13]. The appropriate superfields represent, in one
or another way, the $(4,4)$ $2D$ twisted
multiplet \cite{{IK},{GHR}}.

Much less is known about $(2,2)$ and $(4,4)$ sigma models with
non-commuting
complex structures, despite the fact that most of the
corresponding group manifold WZNW sigma models \cite{belg} fall into this
category
\cite{RSS}. In particular, it is unclear how
to describe them off shell in general. As was argued in Refs.
\cite{{GHR},{BRL},{RSS}}, twisted $(2,2)$ and $(4,4)$ multiplets are
not suitable for this purpose. It has been then suggested to make use
of some other off-shell representations of $(2,2)$ \cite{{BRL},{DS}}
and $(4,4)$ \cite{{BRL},{RIL}} worldsheet SUSY. However, it is an open
question whether the relevant actions correspond to generic sigma models of
this type.

In this talk we describe another approach to the off-shell
description of general $(4,4)$ sigma models with torsion, exploiting an
analogy with general torsionless hyper-K\"ahler $(4,4)$
sigma models in $SU(2)$ harmonic superspace [2 - 4]. The presentation
is based upon two recent preprints of the author \cite{{Iv},{Iv2}} and
his paper with A. Sutulin \cite{IS}.
We start from
a dual form of the general action of $(4,4)$ twisted superfields in
$SU(2)\times SU(2)$ analytic harmonic superspace with two independent
sets of harmonic variables \cite{IS} and construct a direct
$SU(2)\times SU(2)$ harmonic analog of the hyper-K\"ahler $(4,4)$ action.
The form of the action obtained, contrary to the
torsionless case, proves to be severely constrained by the integrability
conditions following from the commutativity of the left and right harmonic
derivatives. While for four-dimensional bosonic manifolds
the resulting action is reduced to that of twisted superfield,
for manifolds of dimension $4n,\;n\geq 2$, the generic action
{\it cannot be written only in terms of twisted superfields}.
Its most characteristic features are (i) the unavoidable presence of
infinite number of
auxiliary fields and (ii) a nonabelain and in general nonlinear gauge
symmetry which ensures the necessary
number of propagating fields. These symmetry and action
are harmonic analogs of the Poisson gauge symmetry and
actions recently discussed in \cite{{Ikeda},{austr}}.
For an interesting subclass of these actions, harmonic analogs of
the Yang-Mills ones, we explicitly
demonstrate that the left and right complex
structures on the bosonic target {\it do not commute}.

\vspace{0.3cm}
\noindent{\bf 2. (4,4) twisted multiplet in SU(2)xSU(2)
harmonic superspace.}
The $SU(2)\times SU(2)$ harmonic superspace is an extension of the
standard real $(4,4)$ $2D$ superspace by two independent sets
of harmonic variables  $u^{\pm 1\;i}$ and $v^{\pm 1\;a}$
($u^{1\;i}u^{-1}_{i} =
v^{1\;a}v^{-1}_{a} = 1$) associated with
the automorphism groups $SU(2)_L$ and $SU(2)_R$ of the left and
right sectors of $(4,4)$ supersymmetry \cite{IS}. The corresponding
analytic subspace is spanned by the following set of coordinates
\be  \label{anal2harm}
(\zeta, u,v) =
(\;x^{++}, x^{--}, \theta^{1,0\;\underline{i}},
\theta^{0,1\;\underline{a}}, u^{\pm1\;i}, v^{\pm1\;a}\;)\;,
\ee
where we omitted the light-cone indices of odd coordinates. The
superscript ``$n,m$'' stands for two independent harmonic $U(1)$ charges,
left ($n$) and right ($m$) ones.

It was argued in \cite{IS} that this type of harmonic superspace is most
appropriate for constructing off-shell formulations of $(4,4)$
sigma models with torsion. This hope mainly relied upon the fact
that the twisted $(4,4)$ multiplet has a natural description as a
real analytic $SU(2)\times SU(2)$
harmonic superfield $q^{1,1}(\zeta,u,v)$ (subjected to some
harmonic constraints). The
most general off-shell action of $n$ such multiplets is
given by the following integral over the analytic superspace
(\ref{anal2harm}) \cite{IS}
\be
S_{q,\omega} = \int \mu^{-2,-2} \{
q^{1,1\;M}(\;D^{2,0} \omega^{-1,1\;M}   +
D^{0,2}\omega^{1,-1\;M} \;) + h^{2,2} (q^{1,1}, u, v) \}\;\;
(M=1,...,n) \;.
\label{dualq}
\ee
where
\bea
D^{2,0} &=& \partial^{2,0} + i\theta^{1,0\;\underline{i}}
\theta^{1,0}_{\underline{i}}\partial_{++}\;,
\;\;D^{0,2} \;=\;\partial^{0,2} + i\theta^{0,1\;\underline{a}}
\theta^{0,1}_{\underline{a}}\partial_{--}
\label{harm2der} \\
(\partial^{2,0} &=& u^{1 \;i}\frac{\partial} {\partial u^{-1 \;i}}\;,\;\;
\partial^{0,2} \;=\; v^{1 \;a}\frac{\partial} {\partial v^{-1 \;a}})
\nonumber
\eea
are the left and right analyticity-preserving harmonic derivatives and
$\mu^{-2,-2}$ is the analytic superspace integration measure. In
(\ref{dualq}) the involved superfields are unconstrained analytic,
so from the beginning the action (\ref{dualq}) contains an infinite number
of auxiliary fields coming from the double harmonic expansions with
respect to the
harmonics $u^{\pm1\;i}, v^{\pm1\;a}$. However, after varying
with respect to the Lagrange
multipliers $\omega^{1,-1\;M}, \omega^{-1,1\;M}$, one comes to the action
written only in terms of $q^{1,1\;N}$ subjected to the harmonic constraints
\be  \label{qconstr}
D^{2,0} q^{1,1\;M} = D^{0,2} q^{1,1\;M} = 0\;.
\ee
For each value of $M$ these constraints define the $(4,4)$ twisted
multiplet in the $SU(2)\times
SU(2)$ harmonic superspace ($8+8$ components off-shell), so the action
(\ref{dualq}) is a dual form of the general off-shell action of $(4,4)$
twisted multiplets \cite{IS}
\be
S_{q} = \int \mu^{-2,-2}\; h^{2,2}(q,u,v)\;.  \label{qact}
\ee
As an important particular example of such a $q^{1,1}$ action we give
the action of $(4,4)$ extension of the group manifold $SU(2)\times U(1)$
WZNW sigma model
\be \label{wzwaction}
S_{wzw} = -\frac{1}{4\kappa ^2} \int \mu^{-2,-2} \;\hat{q}^{1,1}
\hat{q}^{(1,1)}
\left( \frac{1}{(1+X)X} -\frac{\mbox{ln}(1+X)}{X^2} \right)\;.
\label{confact}
\ee
Here
\be
\hat{q}^{1,1} = q^{1,1} - c^{1,1}\;,\;X = c^{-1,-1}\hat{q}^{1,1}\;,\;
c^{\pm 1,\pm 1} = c^{ia}u^{\pm1}_iv^{\pm1}_a \;,\;
c^{ia}c_{ia} = 2\;.
\ee
Despite the presence of an extra quartet constant $c^{ia}$ in the
analytic superfield Lagrangian, the action (\ref{wzwaction})
actually does not depend
on $c^{ia}$ \cite{IS} as it is invariant under arbitrary
rescalings and $SU(2)\times SU(2)$ rotations of this constant.

The crucial feature of the dual action (\ref{dualq}) is the abelian gauge
invariance
\be
\delta \;\omega^{1,-1\;M} = D^{2,0} \sigma^{-1,-1\;M}
\;, \; \delta \;\omega^{-1,1\;M}
= - D^{0,2} \sigma^{-1,-1\;M}
\label{gauge}
\ee
where $\sigma^{-1,-1\;M}$ are iunconstrained analytic
superfield parameters.
This gauge freedom ensures the on-shell equivalence
of the $q,\omega$ formulation of
the twisted multiplet action to its original $q$ formulation
\p{qact} \cite{IS}: it
neutralizes superfluous physical dimension component fields in the
superfields $\omega^{1,-1\;M}$ and $\omega^{-1,1\;M}$ and
thus equalizes the number of propagating fields in both formulations. It
holds already at the free level, with
$h^{2,2}$ quadratic in $q^{1,1\;M}$, so it is natural to expect that
any reasonable generalization of the action (\ref{dualq}) respects
this symmetry or a generalization of it. We will see soon that this is
indeed so.

After identifying harmonics $u$ and $v$ as well as two harmonic $U(1)$
charges, and defining
\bea
D^{(+2)} &\equiv & (D^{2,0} + D^{0,2})|_{u=v}\;, \; \; \; \omega^{N} \equiv
\omega^{1,-1\;N}|_{u=v} = \omega^{-1,1\;N}|_{u=v}\;, \nn \\
l^{(+2)\;N} &\equiv & q^{1,1\;N}|_{u=v}\;, \; \;\;
L^{(+4)} \equiv h^{2,2}|_{u=v}
\eea
the action (\ref{dualq}) is reduced to the dual action of tensor
$(4,4)$ $2D$ multiplet in the $SU(2)$ $2D$ harmonic superspace
\cite{GIO}
\be \label{laction}
S_{l} = \int \mu^{(-4)} \{ -2\omega^{N}D^{(+2)}l^{(+2)\;N} +
\tilde{L}^{(+4)}(l, u) \}\;.
\ee
Varying (\ref{laction}) with respect to
$\omega^N$, we arrive at the action which contains only the
$\tilde{L}^{(+4)}(l, u)$ part,
\be \label{laction1}
S_{l} = \int \mu^{(-4)} \tilde{L}^{(+4)}(l, u) \;,
\ee
with the superfield $l^{(+)\;N}$ subjected to the constraint
\be \label{lconstr}
D^{(+2)}l^{(+2)\;N} = 0\;.
\ee
This is just the harmonic superspace action and constraint of $N=2\;\;4D$
($(4,4)\;\;2D$) tensor multiplet \cite{GIO}. The action \p{laction} is a
particular case of the general action of torsionless hyper-K\"ahler
$(4,4)$ sigma models in the $\omega, l^{(+2)}$ representation
\cite{{GIO},{GIOSap}}. Its specificity consists in that it respects
$n$ independent abelian isometries realized as constant shifts of the
Lagrange multipliers $\omega^N$.

\vspace{0.3cm}
\noindent{\bf 3. More general (4,4) sigma models with torsion.}
The dual twisted multiplet action (\ref{dualq}) is
a good starting point for constructing more general actions which, as we
will show, encompass sigma models with non-commuting left and right
complex structures.

It is useful to apply to the suggestive analogy with the general
action of hyper-K\"ahler
$(4,4)$ sigma models in the $SU(2)$ harmonic superspace \cite{GIKOS}.
This action in the $\omega, l^{+2}$ representation \cite{GIOSap}
looks very similar to (\ref{dualq}), the $SU(2)$ analytic superfield
pair $\omega^M, l^{+2\;M}$ being the clear analog of the
$SU(2)\times SU(2)$
analytic superfield triple $\omega^{1,-1\;M}, \omega^{-1,1\;M},
q^{1,1\;M}$ and the general hyper-K\"ahler potential being analogous to
$h^{2,2}$.
However, this analogy breaks in that the hyper-K\"ahler potential is
in general an arbitrary function of all involved
superfields and harmonics while $h^{2,2}$ in (\ref{dualq})
depends only on $q^{1,1\;M}$ and harmonics. As we just saw, (\ref{dualq}
is the direct analog of the particular class of hyper-K\"ahler
actions \p{laction}, with the hyper-K\"ahler potential displaying
no dependence on $\omega^{M}$. The general hyper-K\"ahler $(4,4)$
action can be obtained from \p{laction} by including an arbitrary
dependence on $\omega^M$ in $L^{(+4)}$. Then an obvious way to
generalize (\ref{dualq}) to cover a wider class of torsionful
$(4,4)$ models is to allow for a dependence on
$\omega^{1,-1\;M}, \omega^{-1,1\;M}$ in $h^{2,2}$.

With these reasonings in mind, we take as an ansatz for the general
action the following one
\be
S_{gen} = \int \mu^{-2,-2} \{
q^{1,1\;M}(\;D^{2,0} \omega^{-1,1\;M}   +
D^{0,2}\omega^{1,-1\;M} \;) + H^{2,2} (q^{1,1}, \omega^{1,-1},
\omega^{-1,1},
u, v) \}\;,
\label{genact}
\ee
where for the moment the $\omega$ dependence in $H^{2,2}$ is not fixed.
In Sect.5 we will show that one can arrive at this action proceeding
>from the most general $q,\omega$ action containing first order harmonic
derivatives. But, for the time being, it is convenient for us to
rely on the analogy with the $(4,4)$ sigma model action in
the $SU(2)$ harmonic superspace.

Now we are approaching the most important point. Namely, we are going to
show that,
contrary to the case of $SU(2)$ harmonic action of torsionless $(4,4)$
sigma models, the $\omega$ dependence of the potential $H^{2,2}$ in
(\ref{genact}) is completely specified
by the integrability conditions following from the
commutativity relation
\be \label{comm}
[\;D^{2,0}, D^{0,2}\;] = 0\;.
\ee

To this end, let us write the equations of motion corresponding to
(\ref{genact})
\bea
D^{2,0}\omega^{-1,1\;M} + D^{0,2}\omega^{1,-1\;M} &=& -
\frac{\partial H^{2,2} (q,\omega,u,v)}{\partial q^{1,1\;M}}\;,
\label{eqom} \\
D^{2,0}q^{1,1\;M} \;=\;
\frac{\partial H^{2,2} (q,\omega,u,v)}{\partial \omega^{-1,1\;M}}\;,
\;\;
D^{0,2}q^{1,1\;M} &=&
\frac{\partial H^{2,2} (q,\omega,u,v)}{\partial \omega^{1,-1\;M}}\;.
\label{eqqu}
\eea
Applying the intgrability condition (\ref{comm}) to the pair of
equations (\ref{eqqu}) and imposing a natural requirement that it
is satisfied as a consequence of the equations of motion (i.e. does not
give rise to any new dynamical restrictions), after some algebra
we arrive at the following set of self-consistency relations
\bea
&& \frac{\partial^2 H^{2,2}}{\partial \omega^{-1,1\;N} \partial
\omega^{-1,1\;M}}
\;=\;
\frac{\partial^2 H^{2,2}}{\partial \omega^{1,-1\;N} \partial
\omega^{1,-1\;M}}
\;=\; \frac{\partial^2 H^{2,2}}{\partial \omega^{1,-1\;(N}
\partial \omega^{-1,1\;M)}}
\;=\; 0\;, \label{consN1} \\
&& \left( \partial^{2,0} + \frac{\partial H^{2,2}}{\partial
\omega^{-1,1\;N}}
\;\frac{\partial}{\partial q^{1,1\;N}}
-{1\over 2}\; \frac{\partial H^{2,2}}{\partial q^{1,1\;N}}
\;\frac{\partial}{\partial \omega^{-1,1\;N}}
\right)
\frac{\partial H^{2,2}}{\partial \omega^{1,-1\;M}} \nonumber \\
&& -\left( \partial^{0,2} + \frac{\partial H^{2,2}}{\partial
\omega^{1,-1\;N}}
\;\frac{\partial}{\partial q^{1,1\;N}}
-{1\over 2}\; \frac{\partial H^{2,2}}{\partial q^{1,1\;N}}
\;\frac{\partial}{\partial \omega^{1,-1\;N}}
\right)
\frac{\partial H^{2,2}}{\partial \omega^{-1,1\;M}} \;=\; 0\;.
\label{consN2}
\eea
Eqs. (\ref{consN1}) imply
\bea
H^{2,2} &=& h^{2,2}(q,u,v) + \omega^{1,-1\;N} h^{1,3\;N}(q,u,v)
+ \omega^{-1,1\;N} h^{3,1\;N}(q,u,v) \nonumber \\
&& + \;\omega^{-1,1\;N}\omega^{1,-1\;M}
h^{2,2\;[N,M]}(q,u,v)\;. \label{Hgen}
\eea
Plugging this expression into the constraint (\ref{consN2}),
we finally deduce four independent constraints on the potentials
$h^{2,2}$, $h^{1,3\;N}$, $h^{3,1\;N}$ and $h^{2,2\;[N,M]}$
\bea
&& \nabla^{2,0} h^{1,3\;N} - \nabla^{0,2} h^{3,1\;N} + h^{2,2\;[N,M]}
\;\frac{\partial h^{2,2}}{\partial q^{1,1\;M}} \;=\; 0 \label{1} \\
&& \nabla^{2,0} h^{2,2\;[N,M]} -
\frac{\partial h^{3,1\;N}}{\partial q^{1,1\;T}} \;h^{2,2\;[T,M]} +
\frac{\partial h^{3,1\;M}}{\partial q^{1,1\;T}}\; h^{2,2\;[T,N]} \;=\; 0
\label{2} \\
&& \nabla^{0,2} h^{2,2\;[N,M]} -
\frac{\partial h^{1,3\;N}}{\partial q^{1,1\;T}}\; h^{2,2\;[T,M]} +
\frac{\partial h^{1,3\;M}}{\partial q^{1,1\;T}}\; h^{2,2\;[T,N]} \;=\; 0
\label{3} \\
&& h^{2,2\;[N,T]}\;\frac{\partial h^{2,2\;[M,L]}}{\partial q^{1,1\;T}} +
h^{2,2\;[L,T]}\;\frac{\partial h^{2,2\;[N,M]}}{\partial q^{1,1\;T}} +
h^{2,2\;[M,T]}\;\frac{\partial h^{2,2\;[L,N]}}{\partial q^{1,1\;T}}
\;=\; 0
\label{4}
\eea
where
\be
\nabla^{2,0} = \partial^{2,0} + h^{3,1\;N}\frac{\partial}{\partial
q^{1,1\;N}}
\;,\;\;
\nabla^{0,2} = \partial^{0,2} + h^{1,3\;N}\frac{\partial}{\partial
q^{1,1\;N}}
\;.
\ee
and $\partial^{2,0}, \partial^{0,2}$ act only on the ``target''
harmonics, i.e. those appearing
explicitly in the potentials.

Thus we have shown that the direct generalization of the generic
hyper-K\"ahler $(4,4)$ sigma model action to the torsionful
case is given by the action
\bea
S_{q,\omega} &=&
\int \mu^{-2,-2} \{\; q^{1,1\;M}D^{0,2}\omega^{1,-1\;M} +
q^{1,1\;M}D^{2,0}\omega^{-1,1\;M} +  \omega^{1,-1\;M}h^{1,3\;M}
\nonumber \\
&&+ \omega^{-1,1\;M}h^{3,1\;M} + \omega^{-1,1\;M} \omega^{1,-1\;N}
\;h^{2,2\;[M,N]} + h^{2,2}\;\}\;, \label{haction}
\eea
where the involved potentials depend only on $q^{1,1\;M}$ and target
harmonics and
satisfy the target space constraints (\ref{1}) - (\ref{4}). These
constraints certainly encode a nontrivial geometry which for the time
being is
unclear to us. To reveal it we need to solve the
constraints, which is still to be done. At present we are only aware of
their particular solution which will be discussed in the next
section.

In the rest of this section we present a set of invariances of
the action (\ref{haction}) and constraints (\ref{1}) - (\ref{4})
which can be useful for understanding the underlying geometry of the
given class of sigma models.

One of these invariances is a mixture of reparametrizations in
the target space (spanned by the involved superfields and target
harmonics) and the transformations which are bi-harmonic
analogs of hyper-K\"ahler
transformations of Refs. \cite{{BGIO},{GIOSap}}.
It is realized by
\bea
\delta q^{1,1\;N} &=& \lambda^{1,1\;N}\;,\;\; \delta \omega^{-1,1\;N}
\;=\;
-\frac{\partial \lambda^{0,2}}{\partial q^{1,1\;N}} -
\frac{\partial \lambda^{1,1\;M}}{\partial q^{1,1\;N}}\;
\omega^{-1,1\;M}\;,
\nonumber \\
\delta \omega^{1,-1\;N} &=&
-\frac{\partial \lambda^{2,0}}{\partial q^{1,1\;N}} -
\frac{\partial \lambda^{1,1\;M}}{\partial q^{1,1\;N}}\;
\omega^{1,-1\;M}\;,
\nonumber \\
\delta h^{2,2} &=& \nabla^{2,0} \lambda^{0,2} + \nabla^{0,2}
\lambda^{2,0}\;,
\nonumber \\
\delta h^{3,1\;M} &=& \nabla^{2,0}\lambda^{1,1\;M} + h^{2,2\;[M,N]}\;
\frac{\partial \lambda^{2,0}}{\partial q^{1,1\;N}} \nonumber \\
\delta h^{1,3\;M} &=& \nabla^{0,2}\lambda^{1,1\;M} - h^{2,2\;[M,N]}\;
\frac{\partial \lambda^{0,2}}{\partial q^{1,1\;N}} \nonumber \\
\delta h^{2,2\;[N,M]} &=&
\frac{\partial \lambda^{1,1\;N}}{\partial q^{1,1\;L}}\; h^{2,2\;[L,M]} -
\frac{\partial \lambda^{1,1\;M}}{\partial q^{1,1\;L}}\; h^{2,2\;[L,N]} \;,
\label{hrep}
\eea
all the involved transformation parameters being unconstrained
functions of $(q^{1,1\;M}, u, v)$. This kind of invariance
can be used to bring the potentials in (\ref{haction}) into
a ``normal'' form similar to
the normal gauge of the hyper-K\"ahler potential (see \cite{GIOSap}).

Much more interesting is another invariance which has no analog in the
hyper-K\"ahler case and is a nonabelain and in general nonlinear
generalization of the abelian gauge invariance (\ref{gauge})
\bea
\delta \omega^{1,-1\;M}  &=&
\left( D^{2,0}\delta^{MN} +
\frac{\partial h^{3,1\;N}}{\partial q^{1,1\;M}}
\right) \sigma^{-1,-1\;N} - \omega^{1,-1\;L} \;
\frac{\partial h^{2,2\;[L,N]}}
{\partial q^{1,1\;M}}\;\sigma^{-1,-1\;N}\;,
\nonumber \\
\delta \omega^{-1,1\;M}  &=&
- \left( D^{0,2}\delta^{MN} + \frac{\partial h^{1,3\;N}}{\partial
q^{1,1\;M}} \right) \sigma^{-1,-1\;N} - \omega^{-1,1\;L} \;
\frac{\partial h^{2,2\;[L,N]}}{\partial q^{1,1\;M}}\;\sigma^{-1,-1\;N}\;,
\nonumber \\
\delta q^{1,1\;M} &=& \sigma^{-1,-1\;N} h^{2,2\;[N,M]}\;. \label{gaugenab}
\eea
As expected, the action is invariant only with taking account of the
integrability
conditions (\ref{1}) - (\ref{4}). In general, these gauge
transformations close with a field-dependent Lie bracket parameter.
Indeed, commuting two
such transformations, say, on $q^{1,1\;N}$, and using the
cyclic constraint (\ref{4}), we find
\be
\delta_{br} q^{1,1\;M} = \sigma^{-1,-1\;N}_{br} h^{2,2\;[N,M]}\;, \;\;
\sigma^{-1,-1\;N}_{br} = -\sigma^{-1,-1\;L}_1 \sigma^{-1,-1\;T}_2
\frac{\partial h^{2,2\;[L,T]}}{\partial q^{1,1\;N}}\;.
\ee
We see that eq. (\ref{4}) guarantees the nonlinear closure of the
algebra of gauge transformations (\ref{gaugenab}) and so it is a group
condition similar to the Jacobi identities.

Curiously enough, the
gauge transformations (\ref{gaugenab}) augmented with the group
condition (\ref{4})
are precise bi-harmonic counterparts of the two-dimensional version of
basic relations of
the Poisson nonlinear gauge theory
which recently received some attention
\cite{{Ikeda},{austr}} (with the evident correspondence
$D^{2,0}, D^{0,2} \leftrightarrow \partial_\mu$;
$\omega^{1,-1\;M}, - \omega^{-1,1\;M} \leftrightarrow
A_\mu^M $; $\mu = 1,2$). The action (\ref{haction})
coincides in appearance with the general (non-topological) action of
Poisson gauge theory \cite{austr}. The manifold $(q,u,v)$ can be
interpreted as a kind of bi-harmonic extension of some Poisson
manifold and the potential
$h^{2,2\;[N,M]}(q,u,v)$ as a tensor field inducing the Poisson
structure on this extension. We find it remarkable that the harmonic
superspace action of
torsionful $(4,4)$ sigma models deduced using an analogy with
hyper-K\"ahler $(4,4)$ sigma models proved to be
a direct harmonic counterpart of the nonlinear gauge theory action
constructed in \cite{{Ikeda},{austr}} by entirely
different reasoning! We believe that this exciting analogy
is a clue to the understanding of the intrinsic geometry of
general $(4,4)$ sigma models with torsion.

To avoid a possible confusion, it is worth mentioning that the theory
considered {\it is not} a supersymmetric extension of any
genuine $2D$ gauge theory:
there are no gauge fields in the multiplet of physical fields.
The only role of gauge invariance (\ref{gaugenab}) seems to consist in
ensuring the correct number of the sigma model physical fields
($4n$ bosonic and $8n$ fermionic ones).

It should be pointed out that it is the presence of the
antisymmetric potential $h^{2,2\;[N,M]}$ that makes the considered case
nontrivial and, in particular, the gauge invariance (\ref{gaugenab})
nonabelian. If
$h^{2,2\;[N,M]}$ is vanishing, the
invariance gets abelian and the constraints (\ref{2}) - (\ref{4})
are identically satisfied, while (\ref{1}) is solved by
\be \label{hzero}
h^{1,3\;M} = \nabla^{0,2}\Sigma^{1,1\;M}(q,u,v),\;\;
h^{3,1\;M} = \nabla^{2,0}\Sigma^{1,1\;M}(q,u,v)\;,
\ee
with $\Sigma^{1,1\;M}$ being an unconstrained prepotential. Then,
using the target space
gauge symmetry (\ref{hrep}), one may entirely gauge away $h^{1,3\;M},
h^{3,1\;M}$, thereby reducing (\ref{haction}) to the dual
action of twisted $(4,4)$ multiplets (\ref{dualq}). In the case of
one triple $q^{1,1}, \omega^{1,-1}, \omega^{-1,1}$ the potential
$h^{2,2\;[N,M]}$ vanishes identically, so the general action
(\ref{genact}) for $n=1$ is actually equivalent to (\ref{dualq}).
Thus only for $n\geq 2$ a new
class of torsionful $(4,4)$ sigma models comes out.
It is easy to see that the action (\ref{haction}) with non-zero
$h^{2,2\;[N,M]}$ {\it does not} admit any duality transformation
to the form with the superfields $q^{1,1\;M}$ only, because
it is impossible to remove the dependence on $\omega^{1,-1\;N},
\omega^{-1,1\;N}$ from the
equations for $q^{1,1\;M}$ by any
local field redefinition with preserving harmonic analyticity. Moreover,
in contradistinction to the constraints (\ref{qconstr}), these
equations are compatible
only with using the equation for $\omega$'s. So,
the obtained system definitely does not admit in general any dual
description in terms of twisted $(4,4)$ superfields. Hence, the
left and right complex structures on the target space can be
non-commuting. In the next section we will explicitly show this
non-commutativity for a particular class of the models in question.

\vspace{0.3cm}
\noindent{\bf 4. Harmonic Yang-Mills sigma models.}
Here we present a particular solution to the constraints
(\ref{1})-(\ref{4}).
We believe that it shares many features of the general solution which
is as yet unknown.

It is given by the following ansatz
\bea
h^{1,3\;N} &=& h^{3,1\;N} \;=\; 0\;; \; h^{2,2} \;=\; h^{2,2}(t,u,v)\;, \;
\;t^{2,2} \;=\; q^{1,1\;M}q^{1,1\;M}\;; \nonumber \\
h^{2,2\;[N,M]} &=& b^{1,1} f^{NML} q^{1,1\;L}\;, \;b^{1,1} \;=\;
b^{ia}u^1_iv^1_a\;, \; b^{ia} = \mbox{const}\;,
\label{solut}
\eea
where the real constants $f^{NML}$ are totally antisymmetric.
The constraints (\ref{1}) - (\ref{3}) are identically satisfied
with this ansatz, while (\ref{4}) is now none other than the
Jacobi identity which tells us that the constants
$f^{NML}$ are structure constants of some real semi-simple Lie
algebra (the minimal possibility is $n=3$, the
corresponding algebra being $so(3)$). Thus the $(4,4)$ sigma models
associated with
the above solution can be interpreted as a kind of Yang-Mills
theories in the harmonic superspace. They provide the direct
nonabelian generalization
of the twisted multiplet sigma models with the action (\ref{dualq}) which
are thus analogs of two-dimensional abelian gauge theory.
The action (\ref{haction}), related equations of motion and
the gauge transformation laws (\ref{gaugenab}) specialized to the
case (\ref{solut}) are as follows
\bea
S^{YM}_{q,\omega} &=&
\int \mu^{-2,-2} \{\; q^{1,1\;M} (\; D^{0,2}\omega^{1,-1\;M} +
D^{2,0}\omega^{-1,1\;M} + b^{1,1} \;\omega^{-1,1\;L} \omega^{1,-1\;N}
f^{LNM}\; ) \nonumber \\
&&+ \;h^{2,2}(q,u,v) \} \label{haction0}
\eea
\bea
&& D^{2,0} \omega^{-1,1\;N} + D^{0,2} \omega^{1,-1\;N} +
b^{1,1}\;\omega^{-1,1\;S} \omega^{1,-1\;T} f^{STN} \;\equiv\; B^{1,1\;N}
\;=\; - \frac{\partial h^{2,2}}{\partial q^{1,1\;N}}\;, \nonumber \\
&& D^{2,0}q^{1,1\;M} + b^{1,1} \;\omega^{1,-1\;N} f^{NML}  q^{1,1\;L}
\;\equiv \; \Delta^{2,0}q^{1,1\;M}\;=\; 0
\nonumber \\
&& D^{0,2}q^{1,1\;M} - b^{1,1} \; \omega^{-1,1\;N}f^{NML}q^{1,1\;L}
\;\equiv \; \Delta^{0,2}q^{1,1\;M}\;=\; 0
\label{heqmo0} \\
&&\delta \omega^{1,-1\;M}  \;=\;
\Delta^{2,0}\sigma^{-1,-1\;M} \;, \; \delta \omega^{-1,1\;M}  \;=\;
- \Delta^{0,2}\sigma^{-1,-1\;M}\;, \nonumber \\
&& \delta q^{1,1\;M} \;=\;
b^{1,1} \;\sigma^{-1,-1\;N} f^{NML} q^{1,1\;L}\;.
\label{gaugenab0}
\eea

These formulas make the analogy with two-dimensional nonabelian
gauge theory almost literal, especially for
\be \label{free}
h^{2,2} = q^{1,1\;M} q^{1,1\;M}\;.
\ee
Under this choice
$$
q^{1,1\;N} = -{1\over 2}\; B^{1,1\;N}
$$
by first of eqs. (\ref{heqmo0}), then two remaining equations are
direct analogs of two-dimensional Yang-Mills equations
\be \label{litan}
\Delta^{2,0} B^{1,1\;N} = \Delta^{0,2} B^{1,1\;N} = 0\;,
\ee
and we recognize (\ref{haction0}) and (\ref{heqmo0}) as the harmonic
counterpart of the first
order formalism of two-dimensional Yang-Mills theory.
In the general case $q^{1,1\;M}$ is a nonlinear function of
$B^{1,1\;N}$, however for $B^{1,1\;N}$ one still has the same equations
(\ref{litan}).

Now it is a simple exercise to see that in checking the integrability
condition (\ref{comm}) one necessarily needs first of eqs.
(\ref{heqmo0})
$$
[\Delta^{2,0}, \Delta^{0,2}]\; q^{1,1\;M} = - b^{1,1}\;B^{1,1\;N}
f^{NML} q^{1,1\;L} = 2b^{1,1}\; \frac{\partial h^{2,2}}{\partial
t^{2,2}}\; q^{1,1\;N}f^{NML} q^{1,1\;L} \equiv 0\;.
$$
At the same time, in the abelian,
twisted multiplet case this condition is
satisfied without any help from the equation obtained by varying
the action (\ref{dualq}) with respect to $q^{1,1\;N}$.
This property reflects the fact that
the class of $(4,4)$ sigma models we have found cannot be described
only in terms of twisted $(4,4)$ multiplets (of course, in general the
gauge group has the structure of a direct product with
abelian factors; the
relevant $q^{1,1}$'s satisfy the linear twisted multiplet constraints
(\ref{qconstr})).

An interesting specific feature of this ``harmonic Yang-Mills theory'' is
the presence of the doubly charged ``coupling constant'' $b^{1,1}$
in all formulas, which is necessary for the correct balance of the
harmonic $U(1)$ charges. Since $b^{1,1} = b^{ia}u^1_iv^1_a$,
we conclude that in the geometry of the considered class
of $(4,4)$ sigma models a very essential role is played by the quartet
constant $b^{ia}$.
When $b^{ia} \rightarrow 0$, the nonabelian structure contracts into the
abelian one and we reproduce the twisted multiplet action (\ref{dualq}).
We shall see soon that $b^{ia}$ measures the ``strength'' of
non-commutativity of the left and right complex structures.

Let us limit ourselves to the simplest case (\ref{free}) and compute the
relevant bosonic sigma model action and complex structures. We will
do this to the first order in physical bosonic fields, which will be
sufficient to show the non-commutativity of complex structures.

We first impose a kind of Wess-Zumino gauge with respect to
the local symmetry (\ref{gaugenab0}). We choose it so as to gauge
away from $\omega^{1,-1\;N}$ as many components as possible,
while keeping $\omega^{-1,1\;N}$ and $q^{1,1\;N}$ arbitrary.
The gauge-fixed form of $\omega^{1,-1\;N}$ is as follows
\be \label{gaugefix}
\omega^{1,-1\;N} (\zeta, u, v) = \theta^{1,0\;\underline{i}} \;
\nu^{0,-1\;N}_{\underline{i}}(\zeta_R,v) + \theta^{1,0}\theta^{1,0}
\;g^{0,-1\;iN}(\zeta_R,v)u^{-1}_i
\ee
with
$$
\{ \zeta_R \} \equiv \{ x^{++},x^{--}, \theta^{0,1\;\underline{a}} \}\;.
$$
Then we substitute (\ref{gaugefix}) into (\ref{haction0})
with $h^{2,2}$ given by (\ref{free}), integrate over $\theta$'s and $u$'s,
eliminate infinite tails of decoupling auxiliary fields and, after this
routine work, find the physical bosons part of the action (\ref{haction0})
as the following integral over $x$ and harmonics $v$
\be   \label{bosact}
S_{bos} = \int d^2 x [dv] \left( {i\over 2}\;g^{0,-1\;iM}(x,v)\;
\partial_{--} q^{0,1\;M}_i (x,v) \right)\;.
\ee
Here the fields $g$ and $q$ are subjected to the harmonic differential
equations
\bea
&& \partial^{0,2} g^{0,-1\;iM} - 2 (b^{ka}v^{1}_a)\;f^{MNL} q^{0,1\;iN}
g^{0,-1\;L}_k
\;=\; 4i \partial_{++} q^{0,1\;iM} \nonumber \\
&& \partial^{0,2} q^{0,1\;iM} - 2f^{MLN} (b^{ka}v^1_a)
\;q^{0,1\;L}_k \; q^{0,1\;iN}
\;=\; 0  \label{eq12}
\eea
and are related to the initial superfields as
$$
q^{1,1\;M}(\zeta,u,v)| = q^{0,1\;iM}(x,v)u^{1}_i + ... \;, \;\;\;
g^{0,-1\;iN}(\zeta_R, v)| = g^{0,-1\;iN}(x,v)\;,
$$
where $|$ means restriction to the $\theta$ independent parts.

To obtain the ultimate form of the action as an integral over
$x^{++}, x^{--}$, we should solve eqs. (\ref{eq12}), substitute the
solution
into (\ref{bosact}) and do the $v$ integration. Here we solve (\ref{eq12})
to the first
non-vanishing order in the physical bosonic  field $q^{ia\;M}(x)$
which appears as the
first component in the $v$ expansion of $q^{0,1\;iM}$
$$
q^{0,1\;iM}(x,v) = q^{ia\;M}(x)v^1_a + ... \;.
$$

Representing
(\ref{bosact}) as
\be \label{bosact1}
S_{bos} = \int d^2x  \left( G^{M\;L}_{ia\;kb} \partial_{++} q^{ia\;M}
\partial_{--} q^{kb\;L} + B^{M\;L}_{ia\;kb} \partial_{++} q^{ia\;M}
\partial_{--} q^{kb\;L} \right)
\ee
where the metric $G$ and the torsion potential $B$ are,
respectively, symmetric
and skew-symmetric with respect to the simultaneous interchange of
the left and right triples of  their indices, we find that to
the first order
\be \label{GB}
G^{M\;L}_{ia\;kb} = \delta^{ML}\epsilon_{ik} \epsilon_{ab} -
{2\over 3} \epsilon_{ik} f^{MLN} b_{l(a} q^{l\;N}_{b)}\;,
\;\;
B^{M\;L}_{ia\;kb}  =  {2\over 3} f^{MLN} [b_{(i a}q^N_{k)b} +
b_{(ib}q^{N}_{k)a}]\;.
\ee
Note that an asymmetry between the indices $ik$ and $ab$ in the metric
is an artefact of our choice of the WZ gauge in the form (\ref{gaugefix}).
One could choose another gauge so that a symmetry between
the above pairs of $SU(2)$ indices is restored. Metrics in different
gauges are related via the target space $q^{ia\;M}$ reparametrizations.

Finally, let us compute, to the first order in $q^{ia\;M}$,
the left and right complex structures associated with the
sigma models at hand. Following the well-known strategy
\cite{{HoPa},{GHR},{DS}}, we need: (i) to partially go on shell by
eliminating the auxiliary
fermionic
fields; (ii) to divide four supersymmetries in every light-cone
sector into
a $N=1$ one realized linearly and a triplet of
nonlinearly realized extra supersymmetries; (iii) to consider the
transformation laws of the physical bosonic fields $q^{ia\;M}$ under
extra supersymmetries. The complex structures can be read off from these
transformation laws.

In our case at the step (i) we should solve some harmonic differential
equations of motion to express an infinite tail of auxiliary fermionic
fields
in terms of the physical ones and the bosonic fields $q^{ia\;M}$.
The step (ii) amounts
to the decomposition of the $(4,0)$ and $(0,4)$ supersymmetry parameters
$\varepsilon^{i\underline{i}}_{-}$ and $\varepsilon^{a\underline{a}}_+$
as
$$
\varepsilon^{i\underline{i}\;+}\equiv \epsilon^{i\underline{i}}
\varepsilon^+ + i \varepsilon^{(i\underline{i})\;+} \;, \;\;
\varepsilon^{a\underline{a}\;-} \equiv \epsilon^{a\underline{a}}
\varepsilon^- + i \varepsilon^{(a\underline{a})\;-}\;,
$$
where we have kept a manifest symmetry only with respect to the diagonal
$SU(2)$ groups
in the full left and right automorphism groups $SO(4)_L$ and $SO(4)_R$.
At the step (iii) we should redefine the physical fermionic fields so
that the singlet supersymmetries with the
parameters $\varepsilon_-$ and $\varepsilon_+$ be realized linearly.
We skip the details and present the final form of the on-shell
supersymmetry transformations of $q^{ia\;M}(x)$
\be
\delta q^{ia\;M} =
\varepsilon^+ \;\psi^{ia\;M}_+ +i \varepsilon^{(kj)\;+}
\;\left( F_{(kj)} \right)^{ia\;M}_{lb\;L}\;\psi^{lb\;L}_+
+
\varepsilon^- \;\chi^{ia\;M}_- + i\varepsilon^{(cd)\;-}
\;\left( \hat{F}_{(cd)} \right)^{ia\;M}_{lb\;L}\;\chi^{lb\;L}_-\;.
\ee
Introducing the matrices
$$
F^n_{(+)} \equiv  (\tau^n)^k_j F^{(j}_{\;\;\;\;k)}\;,\;\;
F^m_{(-)} \equiv  (\tau^m)^c_d \hat{F}^{(d}_{\;\;\;\;c)}\;,
$$
$\tau^n$ being Pauli matrices, we find that in the first order in
$q^{ia\;M}$ and $b^{ia}$
\bea
F^n_{(+)} &=& -i \tau^n \otimes I \otimes I + {i\over 3}\; [\;M_{(+)},
\tau^n \otimes I \otimes I\;]
\nonumber \\
F^n_{(-)} &=& -i I\otimes \tau^n \otimes I + {i\over 3}\;
[\;M_{(-)}, I\otimes \tau^n\otimes I\;] \label{cstr}
\eea
\be
\left( M_{(+)} \right)^{ia\;M}_{kb\;N} = -2\; f^{MLN}\left(
b^{(i}_{b} q^{a\;L}_{\;\;k)} + b^{(i a} q^{L}_{\;\;k)b} \right)\;, \;\;
\left( M_{(-)} \right)^{ia\;M}_{kb\;N} =
2\;f^{MLN} \;b^i_{(b}q^{a)\;L}_k \;,\label{Matr}
\ee
where the matrix factors in the tensor products are arranged so that they
act, respectively,
on the subsets of indices $i,j,k,...$, $a,b,c,...$, $M,N,L,...$.

It is easy to see that the matrices $F^n_{(\pm)}$ to the first order
in $q$, $b$ possess all the standard properties of
complex structures needed for on-shell $(4,4)$ SUSY \cite{{GHR},{HoPa}}.
In particular, they form a
quaternionic algebra
$$
F^n_{(\pm)}F^m_{(\pm)} = - \delta^{nm} + \epsilon^{nms} F^s_{(\pm)}\;,
$$
and satisfy the covariant constancy conditions
$$
{\cal D}_{lc\;K} \left( F^n_{(\pm)} \right)^{ia\;M}_{kb\;N} =
\partial_ {lc\;K}\left( F^n_{(\pm)} \right)^{ia\;M}_{kb\;N} -
\Gamma^{\;\;\;\;\;\;\;\;\;\;\;jd\;T}_{(\pm)\;lc\;K\;\;kb\;N}
\left( F^n_{(\pm)} \right)^{ia\;M}_{jd\;T} +
\Gamma^{\;\;\;\;\;\;\;\;\;\;\;ia\;M}_{(\pm)\;lc\;K\;\;jd\;T}
\left( F^n_{(\pm)} \right)^{jd\;T}_{kb\;N} = 0
$$
with
$$
\Gamma^{\;\;\;\;\;\;\;\;\;\;\;jd\;T}_{(\pm)\;lc\;M\;\;kb\;N} \equiv
\Gamma^{\;\;\;jd\;T}_{lc\;M\;\;kb\;N} \mp
T^{\;\;\;jd\;T}_{lc\;M\;\;kb\;N}\;,
$$
where $\Gamma$ is the standard Riemann connection for the
metric (\ref{GB}) and $T$ is the torsion
$$
T_{ia\;M\;\;kb\;N\;\;ld\;T} = {1\over 2} \left( \partial_{ia\;M}
B^{N\;T}_{kb\;ld} + \;\;cyclic \right)\;.
$$
It is also straightforward to check two remaining
conditions of the on-shell $(4,4)$ supersymmetry (the hermiticity of the
metric with respect to both sets of complex structures and the vanishing
of the related Nijenhuis tensors). In the present case all these
conditions are guaranteed to be fulfilled because we proceeded from a
manifestly $(4,4)$ supersymmetric off-shell superfield formulation.

It remains to find the commutator of complex structures. The
straightforward computation (again, to the first order in fields)
yields
\bea
[\;F^n_{(+)}, F^m_{(-)}\;] &=&
(\tau^n\otimes I\otimes I) M_{(-)}
(I\otimes \tau^m\otimes I) +
(I\otimes \tau^m\otimes I) M_{(-)}
(\tau^n\otimes I\otimes I) \nonumber \\
&-& (\tau^n\otimes
\tau^m\otimes I)
M_{(-)} -
M_{(-)}(\tau^n\otimes
\tau^m\otimes I)
 \;\neq \; 0\;.
\label{commstr}
\eea

Thus in the present case in the bosonic sector we encounter
a more general geometry compared to the one associated with
twisted $(4,4)$ multiplets. The basic characteristic
feature of this geometry is the non-commutativity  of the left and right
complex structures.
It is easy to check this property also for general
potentials $h^{2,2}(q,u,v)$ in (\ref{haction0})\footnote{This implies,
in particular, that a subclass of metrics associated with
twisted $(4,4)$ multiplets,
for dimensions $4n,\;n\geq 3$, admits a deformation
which preserves $(4,4)$ SUSY but makes the left and right complex
structures non-commuting.}. It
seems obvious that the general case (\ref{haction}),
({\ref{1}) - (\ref{4}) reveals the same feature.
Stress once more that this important property
is related in a puzzling way to
the nonabelian structure of the analytic superspace
actions (\ref{haction0}),
(\ref{haction}): the ``coupling constant'' $b^{1,1}$
(or the Poisson potential
$h^{2,2\;[M,N]}$ in the general case) measures the strength of the
non-commutativity of complex structures.

The main purpose of this Section was to explicitly show that in the
$(4,4)$ models we
have constructed the left and right
complex structures on the bosonic target do not commute.
For full understanding of the geometry of these models, at least
in the particular case discussed in this Section,
and for clarifying its relation to the known examples, e.g.,
to the group manifold ones
\cite{belg}, we need the explicit form of the metrics and torsion
potentials in (\ref{bosact1}). This amounts to finding the
complete (non-iterative)
solution to eqs. \p{eq12} and their generalization to the case of
non-trivial potentials $h^{2,2}(t,u,v)$ in \p{haction0}. A work along
this line is now in progress. We wish to point out that one of the merits
of the off-shell formulation proposed lies in the fact that,
like in the case of $(4,4)$ sigma models without torsion \cite{GIOShk} or
$(4,0)$ models \cite{DVal}, we can {\it explicitly}
compute the bosonic metrics starting from the unconstrained superfield
action \p{haction0} (or its generalization corresponding to the general
solution of constraints \p{1} - \p{4}). These metrics are guaranteed to
satisfy all the conditions of on-shell $(4,4)$ supersymmetry listed in
refs. \cite{{GHR},{HoPa}}. It is worth mentioning that the
latter conditions, in their own right, do not
provide us with any explicit recipe for computing the metrics.

Though we are not yet aware of the detailed properties of the
corresponding bosonic metrics (singularities, etc.),
in the particular case \p{solut} we know some of their isometries.
Namely, the action \p{haction0} and its bosonic part \p{bosact1} (for
any choice of $h^{2,2}(t, u, v)$ in \p{solut})
respect invariance under the global transformations of the group
with structure constants $f^{MNL}$. This suggests a link with the group
manifold $(4,4)$ models \cite{belg}.

Our last comment in this Section concerns the relation with the
paper \cite{DS}. Its authors
studied a superfield description of $(2,2)$ sigma models with
non-commuting structures and found a set of nonlinear constraints on the
Lagrangian which somewhat resemble eqs. \p{1} - \p{4}. An
essential difference of their approach from ours seems to
consist in that it does not allow a smooth limiting transition
to the case with commuting structures.

\vspace{0.3cm}
\noindent{\bf 5. The action (13) as a gauge-fixed form of
general $q, \omega$ action.} Here we show that one can come to the
ansatz \p{genact} with constraints \p{consN1}, \p{consN2} starting from
the most general analytic harmonic superspace action of superfields
$q^{1,1\;N}$, $\omega{1,-1\;N}$, $\omega^{-1,1\;N}$ and systematically
using in this action consequences of the general integrability
condition \p{comm} combined with a freedom of target space
reparametrizations.

The most general action linear in the harmonic derivatives of the
involved superfields is given by \cite{Iv}
\bea
S_{q,\omega} &=& \int \mu^{-2,-2} \{ H^{2,2} + H^{-1,1\;M}
D^{2,0}q^{1,1\;M}
+ H^{1,-1\;M}D^{0,2}q^{1,1\;M} + H^{1,1\;M}D^{2,0}\omega^{1,-1\;M}
\nonumber \\
&& + \tilde{H}^{1,1\;M}D^{0,2}\omega^{-1,1\;M}
+ H^{-1,3\;M} D^{2,0}\omega^{1,-1\;M} +
H^{3,-1\;M} D^{0,2}\omega^{-1,1\;M}  \} \nonumber \\
&\equiv & \int \mu^{-2,-2} {\cal L}^{2,2}_{q, \omega}
(q, \omega, u, v)\;,
\label{verygen}
\eea
where {\it a priori} all the potentials $H$ are arbitrary functions
of the superfields $q^{1,1\;M}$, $\omega^{1,-1\;M}$,
$\omega^{-1,1\;M}$ and harmonics $u, v$.

We will try to use the set of target space gauge invariances
of the type inherent to hyper-K\"ahler $(4,4)$ actions
\cite{{BGIO},{GIOSap}} in order to reduce the number of independent
potentials as much as possible.

One type of such invariances of the action (\ref{verygen}) is related
to reparametrizations of the involved superfields
\bea
\delta q^{1,1\;M} &=& \Lambda^{1,1\;M} (q,\omega, u,v)\;,\;\;
\delta \omega^{1,-1\;M} \;=\; \Lambda^{1,-1\;M} (q,\omega, u,v)\;,
\nonumber \\
\delta \omega^{-1,1\;M} &=& \Lambda^{-1,1\;M} (q,\omega, u,v)\;.
\label{rep}
\eea
It is straightforward to find the transformations of the potentials
such that the action is form-invariant. Their explicit structure is
not too enlightening.

Another type of invariance is similar to the hyper-K\"ahler one
\cite{{BGIO},{GIOSap}}
and is related to the freedom of adding full harmonic deriavtives
to the superfield Lagrangian in (\ref{verygen})
\be
{\cal L}^{2,2}_{q,\omega} \Rightarrow {\cal L}^{2,2}_{q,\omega}  +
D^{2,0} \Lambda^{0,2} + D^{0,2} \Lambda^{2,0}\;, \label{analhk}
\ee
$$
\Lambda^{2,0} = \Lambda^{2,0}(q,\omega,u,v) \;,\;\;
\Lambda^{0,2} = \Lambda^{0,2}(q,\omega,u,v) \;.
$$
Once again, it is easy to indicate how the potentials should transform to
generate the shifts (\ref{analhk}). It will be important for our
consideration that, assuming the existence of the flat limit
(given by the action (\ref{dualq}) with
$L^{2,2} (q, u, v) = q^{1,1\;N}q^{1,1\;N}$), the full gauge
freedom (\ref{rep}), (\ref{analhk}) can be fixed so that
\bea
H^{-1,1\;N} &=& \alpha \omega^{-1,1\;N}\;,\;\;
H^{1,-1\;N} \;=\; \beta \omega^{1,-1\;N}\;, \nonumber \\
H^{1,1\;N} &=& (1+\beta)
q^{1,1\;N}\;,\;\; \tilde{H}^{1,1\;N} \;=\; (1+\alpha)q^{1,1\;N} +
\hat{H}^{1,1\;N}\;, \label{gaugef1}
\eea
$\alpha,\;\beta$ being arbitrary parameters. In this gauge the action
still contains four independent potentials,
$H^{2,2}$, $H^{-1,3\;N}$,
$H^{3,-1\;N}$ and $\hat{H}^{1,1\;N}$,
\bea
S_{q,\omega} &=&
\int \mu^{-2,-2} \{ q^{1,1\;M}D^{0,2}\omega^{1,-1\;M} +
(q^{1,1\;M} + \hat{H}^{1,1\;M})D^{2,0}\omega^{-1,1\;M} \nonumber \\
&& + H^{-1,3\;M} D^{2,0}\omega^{1,-1\;M} +
H^{3,-1\;M} D^{0,2}\omega^{-1,1\;M} + H^{2,2} \}\;, \label{verygen2}
\eea
and is invariant under the following target space gauge transformations
which are a mixture of (\ref{rep}) and (\ref{analhk})
(the unconstrained analytic
parameters $\Lambda^{2,0}, \Lambda^{0,2}$ below do not
precisely coincide with those in eq. (\ref{analhk}),
but are related to them in a simple way)
\bea
\delta \hat{H}^{1,1\;M} &=& - \Lambda^{1,1\;M} +
\frac{\partial \Lambda^{0,2}}{\partial \omega^{-1,1\;M}} +
\Lambda^{1,-1\;N}\frac{\partial H^{-1,3\;N}}{\partial
\omega^{-1,1\;M}} +
\Lambda^{-1,1\;N}\frac{\partial \hat{H}^{1,1\;N}}{\partial
\omega^{-1,1\;M}} \nonumber \\
\delta H^{-1,3\;M} &=&
\frac{\partial \Lambda^{0,2}}{\partial \omega^{1,-1\;M}} +
\Lambda^{1,-1\;N}\frac{\partial H^{-1,3\;N}}{\partial
\omega^{1,-1\;M}} +
\Lambda^{-1,1\;N}
\frac{\partial \hat{H}^{1,1\;N}}{\partial
\omega^{1,-1\;M}}
\nonumber \\
\delta H^{3,-1\;M} &=&
\frac{\partial \Lambda^{2,0}}{\partial \omega^{-1,1\;M}} +
\Lambda^{-1,1\;N}\frac{\partial H^{3,-1\;N}}{\partial \omega^{-1,1\;M}}
\nonumber \\
\delta H^{2,2} &=& \partial^{2,0}\Lambda^{0,2} +
\partial^{0,2}\Lambda^{2,0}
+ \Lambda^{1,-1\;N}\partial^{2,0}H^{-1,3\;N} \nonumber \\
&&+ \Lambda^{-1,1\;N}
(\partial^{2,0}\hat{H}^{1,1\;N} +\partial^{0,2}H^{3,-1\;N})
\label{remgauge} \eea
with
\bea
\Lambda^{1,1\;M} &=&
\frac{\partial \Lambda^{2,0}}{\partial \omega^{1,-1\;M}} +
\frac{\partial H^{3,-1\;N}}{\partial \omega^{1,-1\;M}}\;
\Lambda^{-1,1\;N} \nonumber \\
\Lambda^{1,-1\;M} &=& -
\frac{\partial \Lambda^{2,0}}{\partial q^{1,1\;M}} -
\frac{\partial H^{3,-1\;N}}{\partial q^{1,1\;M}} \;
\Lambda^{-1,1\;N} \nonumber \\
\Lambda^{-1,1\;M} &=& - (B^{-1})^{NM}
\left\{ \frac{\partial \Lambda^{0,2}}{\partial q^{1,1\;N}} -
\frac{\partial \Lambda^{2,0}}{\partial q^{1,1\;T}}
\frac{\partial H^{-1,3\;T}}{\partial q^{1,1\;N}} \right\}
\label{reppar} \\
B^{MN} &=& \delta^{MN} +
\frac{\partial \hat{H}^{1,1\;M}}{\partial
q^{1,1\;N}} - \frac{\partial H^{3,-1\;M}}{\partial q^{1,1\;F}}
\frac{\partial H^{-1,3\;F}}{\partial q^{1,1\;N}}\;,\;
B^{MN}(B^{-1})^{NL}
\;=\; \delta^{ML} \nonumber
\eea
(one should add, of course, the coordinate transformations
(\ref{rep}) with the parameters (\ref{reppar})).
Note that in the case of general manifold ($M=1,2...n, n>1$)
it is impossible to gauge away any of the surviving
potentials with the help of this remaining gauge freedom, though one can
still put them in the form similar to the normal gauge of the
hyper-K\"ahler
potential $L^{(+4)}$ \cite{GIOSap}. The fact that there remain
three more potentials besides $H^{2,2}$ (which is a direct analog
of $L^{(+4)}$) is
the essential difference of the considered case with torsion from
the torsionless hyper-K\"ahler case. It is worth mentioning
that upon the reduction to the $(4,4)$ $SU(2)$
harmonic superspace the superfields $\omega^{1,-1\;N}$
and $\omega^{-1,1\;N}$ in (\ref{verygen})
are identified with each other and recognized as the
simgle superfield $\omega^N$, $q^{1,1\;N}
\Rightarrow l^{(+2)\;N}$,  $H^{2,2} \Rightarrow L^{(+4)}$, and the
potentials $\hat{H}^{1,1\;N}$, $H^{-1,3\;N}$, $H^{3,-1\;N}$
are combined into a shift of $l^{(+2)\;N}$. This shift can be
absorbed in an equivalence redefinition of $l^{(+2)\;N}$,
after which one recovers the $\omega, l$ action
of the general $(4,4)$ hyper-K\"ahler sigma model in
some ``flat'' gauge. Note that the potentials in
\p{verygen}, \p{verygen2} will turn out to be severely
constrained, so the reduction just mentioned actually
produces some particular class of hyper-K\"ahler $(4,4)$ actions.

The equations of motion following from the action \p{verygen2} can
be cast in the form
\bea
D^{0,2}\omega^{1,-1\;M} &=&
-\frac{\partial H^{2,2}}{\partial q^{1,1\;M}} -
\left( \delta^{NM} +
\frac{\partial \hat{H}^{1,1\;N}}{\partial q^{1,1\;M}} \right)
D^{2,0}\omega^{-1,1\;N}
\nonumber \\
&&
-  \frac{\partial H^{3,-1\;N}}{\partial q^{1,1\;M}}
D^{0,2}\omega^{-1,1\;N}
- \frac{\partial H^{-1,3\;N}}{\partial q^{1,1\;M}}
D^{2,0}\omega^{1,-1\;N}\;, \label{omeq} \\
D^{0,2}q^{1,1\;M} &=&
T^{1,3\;M} + T^{0,2\;NM}\;D^{2,0}\omega^{-1,1\;N} +
T^{2,0\;NM}\;D^{0,2}\omega^{-1,1\;N} \nn \\
&& + T^{-2,4\;NM}\;D^{2,0}\omega^{1,-1\;N} \equiv J^{1,3\;M}
\label{02qeq} \\
D^{2,0}q^{1,1\;M} &=&
G^{3,1\;M} + G^{2,0\;NM}\;D^{2,0}\omega^{-1,1\;N} +
G^{4,-2\;NM}\;D^{0,2}\omega^{-1,1\;N} \nn \\
&& + G^{0,2\;NM}\;D^{2,0}\omega^{1,-1\;N} \equiv J^{3,1\;M}\;.
\label{20qeq}
\eea
Here, the coefficient functions depend only on the potentials and their
derivatives. It is a straightforward exercise to write down them
explicitly. For simplicity, we do not give these expressions.

The commutativity condition \p{comm} in the present case gives rise to
the following general integrability condition
\be
D^{2,0} J^{1,3\;M} - D^{0,2} J^{3,1\;M} = 0\;, \label{bascons}
\ee
which severely constrains the coefficient functions $T$ and
$G$ in $J^{1,3\;M}$, $J^{3,1\;M}$ and, further, the potentials through
which these functions are expressed. By construction, this condition is
covariant under the target space gauge group \p{remgauge}, \p{reppar}.

To extract the consequences of the integrability
condition (\ref{bascons}), we should explicitly compute the action
of harmonic derivatives on the potentials in $J^{1,3\;N}$, $J^{3,1\;N}$,
use once again the equations of motion \p{omeq} - \p{20qeq} to
eliminate $D^{0,2}q^{1,1\;N}$, $D^{2,0}q^{1,1\;N}$
and $D^{0,2}\omega^{1,-1\;N}$, and finally equate to zero
the coefficients before independent structures in the obtained
equality. These are the unity, the derivatives $D^{2,0}\omega^{1,-1\;M}$,
$D^{0,2}\omega^{-1,1\;M}$,
$D^{2,0}\omega^{-1,1\;M}$, all possible products of these derivatives,
and the second-order derivatives $(D^{2,0})^2\omega^{1,-1\;M}$,
$(D^{0,2})^2\omega^{-1,1\;M}$,  $D^{2,0} D^{0,2}\omega^{-1,1\;M}$,
$(D^{2,0})^2\omega^{-1,1\;M}$. As a result we arrive at the set of
constraints on the potentials $H^{2,2}$, $\hat{H}^{1,1\;N}$,
$H^{-1,3\;N}$
and $H^{3,-1\;N}$. Since
we started from the equations \p{omeq} - \p{20qeq}
which respect the residual target space gauge freedom \p{remgauge},
\p{reppar}, the set of integrability constraints is also covariant.
Some of these constraints are covariant on their own, while others are
mixed under \p{remgauge}. Instead of writing down the full set of
constraints (it looks rather ugly),  we will first discuss a few selected
ones and show that they, being combined with the gauge freedom
\p{remgauge}, \p{reppar},
reduce the number of independent potentials to one $H^{2,2}$ and,
respectively, the action \p{verygen2} to \p{genact}.

As a first step we write down the constraint following from nullifying
the coefficient before $(D^{0,2})^2\omega^{-1,1\;M}$
\be
F^{4,-2\;[M,N]} \equiv
\frac{\partial H^{3,-1\;M}}{\partial \omega^{-1,1\;N}}
+
\frac{\partial H^{3,-1\;M}}{\partial q^{1,1\;S}}
\frac{\partial H^{3,-1\;N}}{\partial \omega^{1,-1\;S}} - \left( M
\leftrightarrow N \right) = 0\;. \label{31cons}
\ee
It is not difficult to verify that this constraint is covariant
with respect to \p{remgauge}, \p{reppar}
\be
\delta  F^{4,-2\;[M,N]}  = \left(
\frac{\partial \Lambda^{-1,1\;T}}{\partial \omega^{-1,1\;M}}
+ \frac{\partial \Lambda^{-1,1\;T}}{\partial q^{1,1\;S}}
\frac{\partial H^{3,-1\;M}}{\partial \omega^{1,-1\;S}} \right)
F^{4,-2\;[T,N]} - \left(M \leftrightarrow N \right)\;.
\label{Ftran}
\ee
Then it immediately follows that $H^{3,-1\;M}$ can be completely
eliminated. Indeed, using gauge freedom  \p{remgauge}, one can
gauge away the totally symmetric parts of all the coefficients in
the Taylor expansion of $H^{3,-1}$ in $\omega^{-1,1\;N}$. The remaining
parts with mixed symmetry are zero in virtue of \p{31cons}.
Thus
\be
H^{3,-1\;M} = 0\;, \label{31zero}
\ee
and the gauge function $\Lambda^{2,0}$ in \p{remgauge}, \p{reppar} gets
restricted in the following way
\be
\frac{\partial \Lambda^{2,0}}{\partial \omega^{-1,1\;M}} = 0 \;\;
\Rightarrow  \Lambda^{2,0} = \Lambda^{2,0} (q^{1,1},
\omega^{1,-1}, u, v)\;.
\label{gauge11}
\ee

With taking account of \p{31zero}, the constraints which follow from
vanishing of the coefficients before
$D^{0,2}D^{2,0} \omega^{-1,1\;N}$,  $(D^{2,0})^2 \omega^{-1,1\;N}$
and $(D^{2,0})^2 \omega^{1,-1\;N}$ in \p{bascons} are, respectively,
of the form
\bea
F^{2,0\;[M, N]} &\equiv &
\frac{\partial \hat{H}^{1,1\;M}}{\partial \omega^{-1,1\;N}} -
\frac{\partial \hat{H}^{1,1\;N}}{\partial \omega^{-1,1\;M}} = 0
\label{11cons} \\
F^{0,2\;[M,N]} &\equiv &
\left(B^{-1}\right)^{MS} \left(
\frac{\partial \hat{H}^{1,1\;S}}{\partial \omega^{1,-1\;N}} -
\frac{\partial H^{-1,3\;N}}{\partial \omega^{-1,1\;S}} \right)
- \left( M \leftrightarrow N \right) = 0 \label{11cons2} \\
F^{-2,4\;[M,N]} &\equiv &
\frac{\partial H^{-1,3\;M}}{\partial \omega^{1,-1\;N}}
-
\frac{\partial H^{-1,3\;N}}{\partial \omega^{1,-1\;M}} = 0\;.
\label{m13cons}
\eea
We will also need the constraint which comes
>from putting to zero the coefficient in front of the product
$\left( D^{2,0}\omega^{1,-1\;N} \right)
\left( D^{0,2}\omega^{-1,1\;K} \right)$
\be
\frac{\partial}{\partial \omega^{-1,1\;K}}
\left\{ \left( B^{-1} \right)^{ML} \left(
\frac{\partial \hat{H}^{1,1\;L}}{\partial \omega^{1,-1\;N}} -
\frac{\partial H^{-1,3\;N}}{\partial \omega^{-1,1\;L}} \right)
\right\} = 0\;.\label{gng1} \ee

The constraint \p{m13cons} together with the gauge freedom associated
with the parameter $\Lambda^{0,2}$ (still unrestricted) allow one to
fully eliminate $H^{-1,3\;M}$
\be
H^{-1,3\;M} = 0\;. \label{13zero}
\ee
Since the expression in the curly brackets in \p{gng1} does not depend on
$\omega^{-1,1\;M}$, and its transformation law starts
with the symmetric inhomogeneous term
$$
- \frac{\partial^2 \Lambda^{2,0}}{\partial \omega^{1,-1\;M}
\partial \omega^{1,-1\;N}}\;,
$$
the part of this expression which is symmetric in the indices $M,N$
can be gauged away. Then
the constraint \p{11cons2} requires the antisymmetric part also
to vanish, whence
\be
\frac{\partial \hat{H}^{1,1\;M}}{\partial \omega^{1,-1\;N}} = 0\;.
\ee

Finally, since $\hat{H}^{1,1\;M}$ does not depend on $\omega^{1,-1\;N}$,
the residual target space gauge freedom supplemented with the
constraint \p{11cons} is still capable to completely
gauge away $\hat{H}^{1,1\;M}$
\be
\hat{H}^{1,1\;M} = 0\;.  \label{11zero}
\ee

As the result of gauge fixings \p{31zero}, \p{13zero} and \p{11zero},
the general action \p{verygen2} is reduced to \p{genact}. The remainder of
consequences of the integrability condition \p{bascons} is reduced to eqs.
\p{consN1}, \p{consN2} already explored.

\vspace{0.3cm}
\noindent{\bf 6. Conclusion.}
To summarize, proceeding from an analogy with
the $SU(2)$ harmonic superspace description of $(4,4)$ hyper-K\"ahler
sigma models, we have constructed off-shell $SU(2)\times SU(2)$
harmonic superspace actions for
a new wide class of $(4,4)$ sigma models with torsion and
non-commuting left and right complex structures on the bosonic target.
The generality of this class has been proven by starting from the
most general analytic superspace action of the analytic superfield
triple $q^{1,1\;N}, \omega^{1,-1\;N}, \omega^{-1,1\;N}$ which is the true
analog of the pair $\omega^N, l^{(+2)\;N}$ of the hyper-K\"ahler case,
and using the target space gauge invariance together with some
consequences of the integrability condition \p{bascons}.

The non-commutativity of target complex structures is directly related to
the remarkable non-abelian Poisson gauge structure of the actions
constructed. One of the most characteristic features of the general
action is the presence of an infinite number of auxiliary fields
and the lacking of dual-equivalent formulations in terms of
$(4,4)$ superfields with finite sets of auxiliary fields.
It would be interesting to see whether
such formulations exist for some particular cases, e.g., those
corresponding to the bosonic manifolds with
isometries. An example of $(4,4)$ sigma model with non-commuting
structures which admits such a formulation has been given in \cite{RIL}.

The obvious problems for further study are to compute the relevant
metrics and torsions in a closed form and to try to utilize the
corresponding
manifolds as backgrounds for
some superstrings. An interesting question is
as to whether the constraints
(\ref{1}) - (\ref{4}) admit solutions corresponding to the $(4,4)$
supersymmetric group
manifold WZNW sigma models. The list of appropriate group manifolds
has been given in \cite{belg}. The lowest dimension manifold
with non-commuting left and right structures \cite{RSS}
is that of $SU(3)$. Its dimension 8 coincides
with the minimal bosonic manifold dimension at which a non-trivial
$h^{2,2\;[M,N]}$ in (\ref{haction}) can appear.

It still remains to prove that the
action (\ref{haction})
indeed describes most general $(4,4)$ models with torsion.
One way to do this is to start, like in the hyper-K\"ahler and
quaternionic cases \cite{{GIOSap},{quat}}, with the constrained
formulation of the relevant geometry in a
real $4n$ dimensional manifold and to reproduce
the potentials in (\ref{haction}) as some fundamental objects
which solve the initial constraints.

We note that the constrained superfield $q^{1,1\;M}$ the
dual action of which was a starting point of our construction,
actually
comprises only one type of $(4,4)$ twisted multiplet \cite{IK}.
There exist other
types which differ in the $SU(2)_L\times SU(2)_R$ assignment of
their components
\cite{{GHR},{GI}}. At present it is unclear how
to simultaneously describe all of them
in the framework of the $SU(2)\times SU(2)$ analytic
harmonic superspace. Perhaps, their actions are related to
those of $q^{1,1}$
by a kind of duality transformation. It may happen, however, that
for their self-consistent description one will need a more general
type of $(4,4)$ harmonic superspace, with the
whole $SO(4)_L\times SO(4)_R$ automorphism group of $(4,4)$ $2D$ SUSY
harmonized. The relevant actions will be certainly more
general than those constructed in \cite{{Iv},{Iv2}}.

\vspace{0.3cm}
\noindent{\bf Acknowledgements.}
A partial support from
the Russian Foundation of Fundamental
Research, grant 93-02-03821, and the International Science Foundation,
grant M9T300, is acknowledged.

\end{document}